\documentclass[12pt,emlines2]{article}
\usepackage{amssymb}
\begin{document}
\setcounter{footnote}{1}
\begin{center}
\hfill ITEP-TH - 33/03 \\
\vspace{1.0in}
{\Large\bf Discrete symmetries of isomonodromic deformations of order two
Fuchsian differential equations}
\end{center}
\bigskip
\begin{center}
{\sc  OBLEZIN Sergei} \\
\end{center}
\begin{center}
{\sf Independent University of Moscow,\\
Moscow Institute of Physics and Technology}\\
{\it e-mail: oblezin@mccme.ru}
\end{center}
\bigskip

\abstract{\footnotesize
In the present work we investigate the group structure of the Schlesinger
transformations for isomonodromic deformations of order two Fuchsian
differential equations. We perform these transformations as
isomorphisms between the moduli spaces of the logarithmic {\it sl}(2)-
connections with fixed eigenvalues of the residues at singular points.
We give a geometrical interpretation of the Schlesinger transformations
and perform our calculations using the techniques of the modifications of
bundles with connections. In order to illustrate the result we present
classical examples of symmetries of the hypergeometric equation, the Heun
equation and the sixth Painlev\'e equation.
}

\begin{center}
\rule{5cm}{1pt}
\end{center}

\bigskip
\setcounter{footnote}{0}

\section{Introduction}

In this paper we discuss discrete transformations of the moduli spaces of
logarithmic {\it sl}(2)-connections with singularities at distinct points
$\{{\it x_{\rm 1},\ldots,x_n}\}$ on the Riemann
sphere $\mathbb{P}^1$ with fixed eigenvalues ($sl(2)$-orbits)
of the residues.
We are interested in the group structure of isomorphisms between such
moduli spaces. Every such connection performes a differential equaton with
regular singularities on $\mathbb{P}^1$ and the eigenvalues correspond to the
local parameters of solutions the equaton.
Our discrete transformations act on
the parameters of the equation and on its solutions preserving their local monodromies.
In other words one can take a fuchsian differential equation of order two
with singularities at $\{{\it x_{\rm 1},\ldots,x_n}\}$ and put it into isomonodromic analytical
family in the folowing way. If $Y(z)$ is the fundamental solution of this
equation then one can take
$$\partial_{\it z}Y(z)\cdot Y(z)^{-1}\,=\,\sum_i\frac{B_i({\it x_{\rm 1},\ldots,x_n})}{\it z - x_i},$$
simultaneously with the condition for $\{B_i\}$
$$dB_i({\it x_1,\ldots,x_n})\,=\,\sum_j[B_j, B_i]\,d\log(x_i-x_j)$$
called the Schlesinger equation; the isomonodromic system is usually called
the Schlesinger system. We consider an initial
data space of such isomonodromic deformation of {\it n} points on
$\mathbb{P}^1$ and the discrete symmetries of this system define a structure of discrete
system with discrete time variables. The initial
data space of Schlesinger system is isomorphic to
the coarse moduli space $\mathcal{M}_{\it n}$ of collections
$(\mathcal{L},\,\nabla,\,\phi;\,\lambda_1,...,\lambda_{\it n} ),$
with a rank 2 bundle $\mathcal{L}$ on $\mathbb{P}^1$, a connection
$\nabla:\mathcal{L}\rightarrow\mathcal{L}\otimes\Omega^1_{\mathbb{P}^1}({\it x_{\rm
1}+...+x_n})$ and  the horizontal isomorphism $\phi : {\it
det}\mathcal{L}\simeq\mathcal{O}_{\mathbb{P}^1}$; the eigenvalues of
the residues of the connection $\nabla$ at
$x_i,\,i=1,\ldots ,n$ are $(\lambda_i,\,-\lambda_i)$.

Our main result is the calculation of the apropriate group of discrete transformations.
For these purposes
we develop a geometric techniques of the modifications ([1]) of vector bundles with connections
in the second and the third sections. In the original work [20] L. Schlesinger
consider these transformations of the systems of isomonodromic deformations; further
the algebraic aspects of Schlesinger transformations are developed
in the paper of M. Jimba and T. Miwa [15] (see also [16]) but without paying attention
to the group structure. Besides, in classical works [20] and [15] they discuss the
monodromy representation in $GL(N)$. In this work we give the beautiful geometric interpretation
of Schlesinger transformations following the work of D. Arinkin and S. Lysenko [1]
and consider the
{\it SL}(2)-case which one can easily generalise to the classical case of {\it GL}(2)-representations;
we investigate the structure of the apropriate group which is more delicate
than in {\it GL}(2)-case. Our basic instrument for calculations is the
technique of modifications of bundles, or F-sheaves.
The original idea of the modifications appeared in the works [10] of Erich
Hecke as correspondances between the spaces of modular forms called the Hecke correspondances.
Further this
construction was applied for the description of moduli spaces of vector
bundles over a curve of abitrary genus in [22]. In the work [5] of V. Drinfeld
the onstructions of F-sheaves for an abitrary global field is presentes; it is
for proving the global Langlands hypothesis for {\it GL}(2) and they are called
Frobenius-Hecke sheaves or "shtukas". Recently modification are widely used in
mathematical physics (see [8], [14], [16], [17], [23]).

The group of discrete transformations of $\mathcal{M}_{\it n}$ is turn out to be isomorphic
to the affine Weyl group of the root system of type ${\it C_n}$:
$${\it W}(\widehat{C}_{\it n})\simeq{\it T}\rtimes
\left(\left(\mathbb{Z}/2\mathbb{Z}\right)^{\it n}\rtimes
\mathfrak{S}_{\it n}\right),$$
where $n$ is the number of singular points ${\it x}_{\it i}$; the group acts
in a way of a braid group.
Let us explain the structure of this group in terms of our problem.
$\mathfrak{S}_{\it n}$ is generated by permutations of the
singular points which present the isomorphisms between the apropriate
initial data spaces
$$\mathcal{M}_{\it n}(\mathcal{L},\nabla; \phi : {\it
det}\mathcal{L}\simeq\mathcal{O}; \lambda_1,...\lambda_i,...,\lambda_j,...,\lambda_{\it n})\,\simeq\,
\mathcal{M}_{\it n}(\mathcal{L},\nabla; \phi : {\it
det}\mathcal{L}\simeq\mathcal{O}; \lambda_1,...,\lambda_j,...,\lambda_i,...,\lambda_{\it n}).$$
The translation part of the group {\it T} acts
by dicrete shifts of eigenvalues of residues of the apropriate connection;
moreover there are short and long shifts. These shifts perform the following
isomorphisms
$$\mathcal{M}_{\it n}(\mathcal{L},\nabla; \phi : {\it
det}\mathcal{L}\simeq\mathcal{O}; \lambda_1,...,\lambda_{\it n})\,\simeq\,
\mathcal{M}_{\it n}(\mathcal{L},\nabla; \phi : {\it
det}\mathcal{L}\simeq\mathcal{O}; \lambda_1,...,\lambda_i+\frac{1}{2},...,\lambda-\frac{1}{2},...,\lambda_{\it n})$$
and
$$\mathcal{M}_{\it n}(\mathcal{L},\nabla; \phi : {\it
det}\mathcal{L}\simeq\mathcal{O}; \lambda_1,...,\lambda_{\it n})\quad\simeq\quad
\mathcal{M}_{\it n}(\mathcal{L},\nabla; \phi : {\it
det}\mathcal{L}\simeq\mathcal{O}; \lambda_1,...,\lambda_k+1,...,\lambda_{\it n})$$
respectively.
Further we can change
the signums of the eigenvalues of the residues at these points;
otherwise we perform a composition of noncorrelated 'local' Weyl
transpositions at the {\it sl}(2)-orbits of the residues of the connection at $x_i$:
$$
\sigma^i:\quad
\left(\begin{array}{cc}
\lambda_i & 0\\
0 & -\lambda_i
\end{array}\right)\longrightarrow
\left(\begin{array}{cc}
-\lambda_i & 0\\
0 & \lambda_i
\end{array}\right);$$
such transformations
generate the normal subgroup $\left(\mathbb{Z}/2\mathbb{Z}\right)^{\it n}$
of the finite group $W(C_n)$ and for
$(\epsilon_{\rm 1},...,\epsilon_n)\in\left(\mathbb{Z}/2\mathbb{Z}\right)^{\it
n}$ the apropriate isomorphism is
$$\mathcal{M}_{\it n}(\mathcal{L},\nabla; \phi : {\it
det}\mathcal{L}\simeq\mathcal{O}; \lambda_1,...,\lambda_{\it n})\quad\simeq\quad
\mathcal{M}_{\it n}(\mathcal{L},\nabla; \phi : {\it
det}\mathcal{L}\simeq\mathcal{O}; \epsilon_1\lambda_1,...,\epsilon_n\lambda_{\it n}).$$
We can combine such transpositions with the permutations of the points and
also get the element of order four. Besides,
if we combine a pair of modifications with the such local transpositions
we get the finite reflection; it describes the structure of the semidirect
product of our group.
From the other hand the translation part is isomorphic to the integral lattice,
generated by the $${\it C_n}\,=\,\langle\pm 2\epsilon_i,\,\pm\epsilon_i\pm\epsilon_j\rangle$$
in the standart basis of $\mathbb{R}^n$ and our group of transformations is isomorphic to the
automorphism group of this lattice.

In order to clarify the above result we analyse some examples in the 4th, 5th and the 6th sections;
they are three classical examples of hypergeometric equation ([2], [14]), Heun's equation
([11], [2], see also [6], [7]) and the sixth
Painlev\'e equation ([19], [1], [18]). This part of the paper also has the meaning of
bibliographical review of classical works on Fuchsian differential equations
with three and four singularities. Cases of more than four points are much
more complicated for analytical calculations and the apropriate groups of
symmetries were almost unstuded in classical literature.

In the case of three regular singularities a fuchsian differential equation of order two on
$\mathbb{P}^1$ is equivalent to the hypergeometric equation. Its
${\it W}(\widehat{\it C}_3)$ discrete
symmetries were studied by K. Gau$\ss$ and E. Kummer; our presentation of
this subject corresponds to [2].

Next step after the hypergeometric equation is the Fuchsian differential
equation of order two with four singularities called the Heun equation ([2]); our
presentation follows the original work [11] of Karl Heun. The apropriate
calculations in the classical sense of K. Gau$\ss$ and E.
Kummer were thoroughly made by K. Heun; these results are performed
in [11] in the form of 192 Heun's relations analogous to the 24
Kummer relations of hypergeometric functions and we present these results in the fifth section.

One can write the Schlesinger equation for the coefficients of Heun's
equation and
in the special case of the Schlesinger system with four regular singularities with
{\it sl}(2)-monodromies get sixth Painlev\'e equation
${\it P}_{\it VI}$. Discrete symmetries of ${\it P}_{\it VI}$ were studied
in [19], [18], [12], [1] and the group is ${\it W}(\widehat{\it F}_4)$ which is isomorphic
to the extension of the Weyl group ${\it W}(\widehat{\it D}_4)$ with the group $\mathfrak{S}_3$
of automorphisms of Dynkin graph ${\it D}_4$.

I'm deeply grateful to A. M. Levin for stating problems and
numerous stimulating discussions.
I want to thank M. A. Olshanetsky for the attention to this work and
for useful advices and remarks.
I'm also thankful to V. Poberezhny for discussions of singular gauge
transformations which are put in section 4.

This work is partially supported by the program for support of the
scientific schools 00-15-96557 and by the grant RFBR 01-01-00539.

\section{Modifications of rank N bundles with connections}

Let $\mathcal{L}$ be a rank N bundle on $\mathbb{P}^{1}$ with a
connection $\nabla$ and suppose ${\it x} \in \mathbb{P}^{1}$. Denote
${\it V}:=\mathcal{L}_{\it x}$ and let ${\it U}\subset{\it V}$ be a
{\it k}-dimensional subspace. Let us not differ $\mathcal{L}$
and the sheaf of its sections and consider the following modifications of
$\mathcal{L}$

$$({\it x},{\it U})^{\it low}(\mathcal{L}):=\{{\it s}\in\mathcal{L} ~|
~{\it s(x)}~\in{\it U}\},$$
$$({\it x, U})^{\it up}(\mathcal{L}):=({\it x, U})^{\it low}(\mathcal{L})\otimes\mathcal{O}({\it x})$$
which are called the lower and the upper modification respectively.
Let us denote the lower modification by $\widetilde{\mathcal{L}}:=({\it x},{\it U})^{\it low}(\mathcal{L})$
and consider the natural map $\widetilde{\mathcal{L}}_{\it x}\longrightarrow\mathcal{L}_{\it
x}$; its image is {\it U}. Put $\widetilde{U}:=
{\it ker}(\widetilde{\mathcal{L}}_{\it x}\longrightarrow\mathcal{L}_{\it x})$ then
$$({\it x}, \widetilde{U})^{\it up}\widetilde{\mathcal{L}}~=~\mathcal{L}.$$
However, the lower and the upper modifications imply the following exact sequences\\
$${\rm 0}\longrightarrow({\it x, U})^{\it low}(\mathcal{L})
\longrightarrow\mathcal{L}\longrightarrow\delta_{\it x}\otimes\mathcal{L}_{\it x}/{\it U}
\longrightarrow ~{\rm 0},$$
$${\rm 0}\longrightarrow\mathcal{L}\longrightarrow({\it x, U})^{\it up}\mathcal{L}
\longrightarrow\delta_{\it x}\otimes{\it U}\otimes\mathcal{O}({\it x})|_{\it x}\longrightarrow{\rm 0}$$
respectively, where $\delta_{\it x}$ is a sky-scraper sheaf with the support at {\it x}.
\\
Roughly speaking, if we have local decomposition ${\it V}\,=\,{\it U}\bigoplus\widetilde{\it U}$
of $\mathcal{L}\simeq{\it V}\otimes\mathcal{O}$ then
$$({\it x, U})^{\it low}(\mathcal{L})={\it U}\otimes\mathcal{O}\bigoplus
\widetilde{U}\otimes\mathcal{O}({\it -x}),$$
$$({\it x, U})^{\it up}(\mathcal{L})={\it U}\otimes\mathcal{O}({\it x})\bigoplus
\widetilde{U}\otimes\mathcal{O}.$$
\\
In other words we change our bundle rescalling the basis of
sections in the neighbourhood of a point {\it x}; if the local basis is
$$\{{\it s}_1({\it z})\,,\ldots,\,{\it s}_{\it N}({\it z})\}$$
$$\mbox{with}\quad{\it U}\otimes\mathcal{O}\simeq
\{{\it s}_1({\it z})\,,\ldots,\,{\it s}_{\it k}({\it
z})\}\quad\mbox{and}\quad
{\widetilde{\it U}}\otimes\mathcal{O}\simeq
\{{\it s}_{k+1}({\it z})\,,\ldots,\,{\it s}_{\it N}({\it z})\}$$
then
the basis of the lower modification of the bundle is generated by the
sections
$$\{{\it s}_1(z),...,{\it s}_{\it k}(z),({\it z-x})\,
{\it s}_{{\it k}+1}(z),...,({\it z-x})\,{\it s}_{\it N}(z)\},$$
and of the upper one by
$$\{({\it z-x})^{-1}\,{\it s}_1(z),...,({\it z-x})^{-1}\,{\it s}_{\it k}(z),
{\it s}_{{\it k}+1}(z),...,{\it s}_{\it N}(z)\}.$$
Consequently, in the punctured neighbourhood we may rewrite the action of modifications
with the following glueing matrices.
$$({\it x, U})^{\it low}=
\left(\begin{array}{cc}
 {\bf 1}_{\it k} & 0 \\
0 & ({\it z-x})\cdot{\bf 1}_{\it N-k}
\end{array}\right),\qquad
({\it x, U})^{\it up}=
\left(\begin{array}{cc}
({\it z-x})^{-1}{\bf 1}_{\it k} & 0 \\
0 & {\bf 1}_{\it N-k}
\end{array}\right),$$ where ${\bf 1}_{\it m}$ is the identity
$({\it m\times m})$-matrix.
Matrix presentation of the
modifications is supposed to be quite obvious and further we widely use it.

Now we discuss the action of modifications on a connection with logarithmic singularities
on $\mathbb{P}^1$ and we need the following
\\
{\bf Definition.} We say $\mathfrak{M}$ is a module with a support {\it S} on the algebraic curve
{\it X} if we have a finite set $S=\{\it x_{\rm 1},...,x_n\}\subset X$ and a positive
integer ${\it n_i}$ for each point ${\it x_i}$ from {\it S}. Sometimes we identify a module with the apropriate
effective divisor $\sum n_i\cdot x_i\,$; for our purposes we consider the module
$$\mathfrak{M}\,=\,\sum\,{\it x_i}.$$
\\
Let us take a look how the modifications change the connection.
Suppose we start from
some connection $\nabla$ on $\mathcal{L}$ and
$$\nabla:\,\mathcal{L}\longrightarrow\mathcal{L}\otimes\Omega^{\rm 1}(\mathfrak{M});$$
that means that $\nabla$ has {\it simple poles} at the support {\it S} of the
module $\mathfrak{M}$.
Let ${\it x\in S}$ be a singular point
of $\nabla$ and ${\it U}\subseteq{\it V}$ is a ${\it Res}_{\it x}\nabla$
-invariant subspace, i. e. $({\it Res_x}\nabla)({\it U)\subseteq U}$ and let us modify the bundle
in this subspace.
At first note that the lower and the upper
modifications at any point ${\it x}~\in\mathbb{P}^{1}$ change the determinant
$${\it det}({\it x, U})^{\it low}\mathcal{L} =
{\it det}\mathcal{L}\otimes\mathcal{O}(-x\cdot{\it dim\,V/U}),\quad
{\it det}({\it x, U})^{\it up}\mathcal{L} ={\it det}\mathcal{L}\otimes\mathcal{O}(x\cdot{\it dim\,U}).$$
For example consider the lower modification $\widetilde{\mathcal{L}}$ with the connection
$$\nabla': \widetilde{\mathcal{L}}
\stackrel{\nabla|_{\widetilde{\mathcal{L}}}}{\longrightarrow}
\mathcal{L}\otimes\Omega(\mathfrak{M})
\stackrel{{\it pr}}{\longrightarrow}
\widetilde{\mathcal{L}}\otimes\Omega(\mathfrak{M})$$
on $\widetilde{\mathcal{L}}$ and on the determinant bundle we get the
connection
$$tr\nabla' = tr\nabla +\frac{dz}{z-x}\cdot{\it dim\,U}.$$

It is principal that we modify the pairs $(\mathcal{L},\,\nabla)$ in
(${\it Res}_{\it x}\nabla$)-invariant subspaces of ${\it V}\subseteq\mathcal{L}_{\it x}$;
otherwise we raise the order of a pole of the connection. In fact, using the
matrix presentation let us wirte the action of the modification of the bundle in
a noninvatiant subspace at $x=0$:
$$
\left(\begin{array}{cccc}
1 & &  0 \\
\\
0 & & {\it z}
\end{array}\right)
\left[{\it d}\quad +\quad
\left(\begin{array}{cc}
{\displaystyle
\frac{\lambda}{\it z}} & {\displaystyle\frac{\epsilon}{\it z}} \\
\\
0 & {\displaystyle-\frac{\lambda}{\it z}}
\end{array}\right)
\right]
\left(\begin{array}{cc}
1 & 0 \\
\\
0 & {\displaystyle\frac{1}{\it z}}
\end{array}\right)
~=~{\it d} +
\left(\begin{array}{cc}
{\displaystyle
\frac{\lambda}{\it z}} & {\displaystyle\frac{\epsilon}{{\it z}^2}} \\
\\
0 & {\displaystyle-\frac{\lambda+1}{\it z}}
\end{array}\right),$$ where {\it z} is a local parameter.
\\
Further we consider a rank 2 bundle $\mathcal{L}$ with ${\it sl}$(2)-connection
$\nabla$ hence we suppose that {\it dim\,U}=1. Morover, we shall perform pairs of the
lower and the upper modifications at
points $x_i$ and $x_j$ respectively to get the bundle $\mathcal{L}''$ with
the same determinant
$${\it det}\mathcal{L}''={\it det}\mathcal{L}\otimes\mathcal{O}({\it x}_{\it j} - {\it x}_{\it i})
\simeq{\it det}\mathcal{L};$$
for this purpose we have to fix a set of compatible isomorphisms
$\mathcal{O}\simeq\mathcal{O}({\it x_i-x_j})$ such that
$$\mathcal{O}\simeq\mathcal{O}({\it x}_{\it i}-{\it x}_{\it j})
\otimes\mathcal{O}({\it x}_{\it j}-{\it x}_{\it k})\simeq
\mathcal{O}({\it x}_{\it i}-{\it x}_{\it k}).$$
\\
Nevertheless if we start from a {\it sl}(2)-connection $\nabla$ then after such
procedure we get the connection
$$\nabla'' = \nabla + {\it P}_{\it U_i}\frac{dz}{z - x_i} -
{\it P}_{\widetilde{U}_{\it j}}\frac{dz}{z - x_j},$$
where ${\it P}_*$ are the projections on the apropriate
{\it Res}$\nabla$-invariant subspaces; it is the {\it gl}(2)-connection.
In order to return to the {\it sl}(2)-connection again we have to add the suitable
1-form
$$\widetilde{\nabla}'' = \nabla'' + \frac{1}{2}\left(\frac{dz}{z - x_j} -
\frac{dz}{z - x_i}\right).$$
\\
This construction performs the nontrivial transformation of
{\it sl}(2)-Schlesinger system of fuchsian type; henceforward it will be one of ours basic
instruments. Precise statements and explanations we peform in next section.

\section{{\it sl}(2)-connections with singularities on $\mathbb{P}^{1}$}

Let us describe our initial data following [1]. Fix a collection $\{\lambda_{\alpha}\}_{\alpha=1}^{\it n}$
of numbers and the module $\mathfrak{M}$ with the support {\it S} at distinct points
$\{x_{\rm 1},...,x_n\}$ on $\mathbb{P}^1$.
There is a three-dimentional group of linear transformations acting on
$\mathbb{P}^1$ so let us suppose that number of the points $n\geq 3$.
Suppose $\mathcal{L}$ be a rank 2 bundle on $\mathbb{P}^{1}$ with fixed
horizontal isomorphism $\phi:\bigwedge^{2}\mathcal{L}\simeq\mathcal{O}$ and with a
connection $\nabla$ with singularities at $\mathfrak{M}=\sum{\it x_i}$; eigenvalues
of {\it Res}$_{\it x_i}\nabla$ are ($\lambda$, -$\lambda$).
Besides, put the following eigenvalue-condition
$$\sum\epsilon_{\it i}\lambda_{\it i}\notin\mathbb{Z},\qquad
(\epsilon_1, \epsilon_2, \epsilon_3, \epsilon_4)\in(\mathbb{Z}/{\rm 2}\mathbb{Z})^4,$$
which guarants the irreducibility of the pair "bundle $\mathcal{L}$ with the
connection $\nabla$" and implies the stability of this pair.
Hence, we fix ${\it sl}$(2)-orbits of residues of the connection.
Denote the eigenspaces of {\it Res}$_{\it x_i}\nabla$
$$\ell_i^{\pm}:={\it ker}({\it Res}_{{\it x}_i}\nabla \mp \lambda _i).$$
\\
For two points ${\it x_i}, {\it x_j}\in S$ consider the modified {\it SL}(2)-bundle
$$\mathcal{L}'' = ({\it x}_{\it j}, \ell_{\it j}^{+})^{\it up}\circ
({\it x}_{\it i}, \ell_{\it i}^{-})^{\it low}\mathcal{L}$$
with modified logarithmic connection $\nabla''$ defined above.
Define an operation $(\downarrow\uparrow)_{\it ij}$ on pairs of
{\it SL}(2)-bundles with connections
$$(\downarrow\uparrow)_{\it ij}:\quad(\mathcal{L}, \nabla)\quad\longmapsto\quad(\mathcal{L}'', \nabla'' +
\omega_{\it ij}),\qquad\omega_{ij} = \frac{1}{2}\left(\frac{dz}{z - x_j} - \frac{dz}{z - x_i}\right).$$

More precisely, we get a nontrivial transformations of the coarse moduli space $\mathcal{M}_{\it n}$
of rank 2 bundles with fixed horizontal isomorphism and logarithmic connection
with fixed eigenvalues of residues on $\mathbb{P}^{1}$.
Let us calculate the correspondance between the eigenvalues under the above isomorphism between
such moduli spaces with different eigenvalues of the residues; precise statement is the
following.
\\
{\bf Proposal.} Modified pair ($\mathcal{L}'', \widetilde{\nabla''}$) is an element
of the coarse moduli space
$\mathcal{M}_{\it n}$. The eigenvalues of
{\it Res}$_{{\it x}_\alpha}\widetilde{\nabla''}$ are
$$\{\lambda_1,\ldots,\lambda_{\it i}+\frac{1}{2},\ldots,
\lambda_{\it j}-\frac{1}{2},\ldots,\lambda_{\it n}\}$$ for the case of a
pair of modifications at distinct points $x_i, x_j\in{\it S}$; if a pair of
modifications is at one point $x_k\in{\it S}$ then the eigenvalues are
$$\{\lambda_1,\ldots,\lambda_{\it k}+1,\ldots,\lambda_{\it n}\}.$$
\\
{\it Proof.} The first part of the statement have been proved. Let us compare
the eigenvalues of the residues of $\nabla$ and modified connection $\nabla''$.
$$
\left\{\begin{array}{ccc}
                \lambda_{\it i} & \lambda_{\it j} \\
                -\lambda_{\it i} & -\lambda_{\it j}
              \end{array}\right\}
\stackrel{\scriptscriptstyle({\it x_i},\ell_{\it i}^-)^{\it low}
\circ({\it x_j}, \ell_{\it j}^+)^{\it up}}{\longrightarrow}
\left\{\begin{array}{ccc}
                \lambda_{\it i}+1 & \lambda_{\it j}-1 \\
                -\lambda_{\it i} & -\lambda_{\it j}
              \end{array}\right\}
\quad\stackrel{+\omega_{ij}}{\longrightarrow}\quad
\left\{\begin{array}{ccc}
                \lambda_{\it i}+1-\frac{1}{2} & \lambda_{\it j}-1+\frac{1}{2}\\
                -\lambda_{\it i}-\frac{1}{2} & -\lambda_{\it j}+\frac{1}{2}
              \end{array}\right\}.
$$
therefore we get the shifts of eigenvalues
$$\lambda_{\it i}\longrightarrow\lambda_{\it i}+\frac{1}{2},
\quad\lambda_{\it j}\longrightarrow\lambda_{\it j}-\frac{1}{2}.$$
In the case of modifications at one point ${\it x_k}$
$$({\it x_i},\ell_{\it i}^-)^{\it low}
\circ({\it x_i}, \ell_{\it i}^-)^{\it up}\,:\quad
\left\{\begin{array}{c}
                \lambda_{\it k}\\
                -\lambda_{\it k}
              \end{array}\right\}
\longrightarrow
\left\{\begin{array}{ccc}
                \lambda_{\it k}+1\\
                -\lambda_{\it k}-1
              \end{array}\right\}$$
we have zero 1-form $\omega_{kk}$ hence the shift is long:
$$\lambda_{\it k}\longrightarrow\lambda_{\it k}+1\qquad\blacksquare$$

Such modifications of pairs $(\mathcal{L},\nabla)$ are the infinite-order
(affine) elements of the apropriate group of transformations of $\mathcal{M}_{\it n}$;
and we see that
the translation part contains both short and long shifts.
Besides this affine symmetry we have evident finite symmetries that are the
permutations of singular points $\{{\it x}_{\alpha}\}\in{\it S}$ and local Weyl
transpositions.
\\
It is significant that our group of transformations is generated by the
elements of order two i.e. reflections. There is a nice classification of
such groups found by H. S. M. Coxeter ([4]) and it alleviates
the description of our group.
Let us arrange the notations. Denote $(i\,j)$ the permutation of distinct
points $x_i, x_j$ of support {\it S} of module $\mathfrak{M}$; moreover we
have the local transpositions
$$\sigma^{\it i}~:\quad
\left(\begin{array}{cc}
\lambda_i & 0\\
0 & -\lambda_i
\end{array}\right)\longrightarrow
\left(\begin{array}{cc}
-\lambda_i & 0\\
0 & \lambda_i
\end{array}\right)$$
from the Weyl group $W(SL(2))$ at each point $x_i\in{\it S}$.
We encode our group of transformations of moduli space $\mathcal{M}_n$ with
the {\bf Coxeter graph} of type $\widehat{C}_n$ using the Coxeter classification.
Finite part of the group generates by the reflections; we encode the
generator with the vertice of the graph and the edges of the graph
correspond to the relations between the generators in the following way.
\\
1. Permutational part $\mathfrak{S}_n$ generates by the $(i\,j)$; the
composition of two neighboured transpositions $(i\,j)(j\,k)\,=\,(i\,j\,k)$ is
the element of order three and we denote this relation with the graph
$(i\,j)\longleftrightarrow (j\,k)$. If a composition of two transpositions has
the order two then we don't put an edge between them.
\\
2. $\left(\mathbb{Z}/{\rm 2}\mathbb{Z}\right)^{\it n}$ generates by the
local Weyl elements $\sigma^i$ and the composition $\sigma^{\it i}\circ(i\,j)$
is the element of order four; we encode this relation withe Coxeter graph
$\sigma^{\it i}\Longleftrightarrow (i\,j)$.
\\
3. Translational part {\it T} generates by the pairs of modifications at
distinct points or at one point and there are short and long affine shifts
respectively.
\\
Henceforward we can give the folowing
\\
{\bf Statement.}
We can present our group with the following  affine
$\widehat{C}_{\it n}$ Coxeter graph.
$$(\uparrow\downarrow)_1\circ\sigma^1\Longrightarrow
(\uparrow\downarrow)_{12}\circ(1\,2)
\longleftrightarrow
(\uparrow\downarrow)_{23}\circ(2\,3)
\longleftrightarrow\ldots\longleftrightarrow
(\uparrow\downarrow)_{n-1,n}\circ(n-1\,n)
\Longleftarrow
(\uparrow\downarrow)_{\it n}\circ\sigma^{\it n}$$

In further sections we analyse this result considering special cases of the
module $\mathfrak{M}=x_1\,+\,x_2\,+\,x_3$ and
$\mathfrak{M}=x_1\,+\,x_2\,+\,x_3\,+\,x_4$. These cases are correspond to the
hypergeometric differential equation and Heun's equation; they were studied in
classical works of K. Gau$\ss$, E. Kummer and K. Heun and correspond to the case of
$gl(2)$-connecitons.
Let us note that the above calculations can be easily generalised to the
$gl(2)$-case. The isomorphisms between the moduli spaces of the pairs
$(\mathcal{L},\nabla)$ with fixed eigenvalues of the residues
will be presented by the same transformations but without the 1-form
$\omega_{ij}$ and all the shifts of eigenvalues of $Res_{x_i}\nabla$ are long. If the eigenvalues
of the connection at a singular points are $(\mu_i, \nu_i)$ and $\widetilde{\mathcal{M}_n}$
denotes the apropriate moduli space of $gl(2)$-connections then the isomorphisms
are
$$\widetilde{\mathcal{M}_{\it n}}(\mathcal{L},\nabla; (\mu_1, \nu_1),...,(\mu_{\it n}, \nu_{\it n}))\quad\simeq\quad
\widetilde{\mathcal{M}_{\it n}}(\mathcal{L},\nabla; (\mu_1, \nu_1),...,(\mu_i\pm 1, \nu_i),...,(\mu_{\it n}, \nu_{\it n})),$$
$$\widetilde{\mathcal{M}_{\it n}}(\mathcal{L},\nabla; (\mu_1, \nu_1),...,(\mu_i, \nu_i),...,(\mu_{\it n}, \nu_{\it n}))\quad\simeq\quad
\widetilde{\mathcal{M}_{\it n}}(\mathcal{L},\nabla; (\mu_1, \nu_1),...,(\nu_i, \mu_i),...,(\mu_{\it n}, \nu_{\it n})),$$
$$\widetilde{\mathcal{M}_{\it n}}(\mathcal{L},\nabla; (\mu_1, \nu_1),...,(\mu_i, \nu_i),...,(\mu_j, \nu_j),...,(\mu_{\it n}, \nu_{\it
n}))\simeq$$
$$\simeq\widetilde{\mathcal{M}_{\it n}}(\mathcal{L},\nabla; (\mu_1, \nu_1),...,(\mu_j, \nu_j),...,(\mu_i, \nu_i),...,(\mu_{\it n}, \nu_{\it n}))$$
Also note that we perform the last permutational isomorphism of the singular points by
the group of linear transformations on the Riemann sphere; this group is
three-dimentional and it is convinient to use this representation for
permuting three and four points on the Riemann sphere.

\section{Classical example: W($\widehat{\it C}_3$)-symmetries\\ of the hypergeometric equation}

At first let us remind the interplay between the meromorphic
connections on $\mathbb{P}^1$ and general differential equations with
singularities.
Let us illustrate our geometric construction in terms of systems of
differential equations in the sense of the fundamental work [3] of A. Bolibruch.
Consider a vector bundle $\mathcal{E}$ and the covariant derivative $\nabla_{\it v}$ for some vector field
{\it v} on $\mathbb{P}^1$ with zeroes at $\mathfrak{M}$; then on
${\it U}=\mathbb{P}^1-{\it S}$ we can trivialize our bundle and hence
$$\nabla_{\it v}|_{\it U}\simeq\partial_{\it v}.$$ For any singular point $x_i\in{\it S}$ we
take its neighbourhood $ V_i$ and we also have a trivialization of our
bundle and use duality between vector fields and
1-forms. Finally we have $\nabla_{\it v}=\partial_{\it v}-\omega$, where $\omega$
is the apropriate 1-form, defined by {\it v}.
If the singularities of $\omega$ are only simple poles then the
trivializing maps at ${\it V_i}$ will be the residues and one can express
$$\nabla=\frac{\it d}{\it dz}\,-
\,\sum_{\it i}\frac{\it res_{x_i}\omega}{\it z\,-\,x_i}.$$
A section $h(z)$ of the {\it GL}(2)-bundle $\mathcal{E}$ is called horizontal with
respect to $\nabla_{\it v}$ if $\nabla_{\it v}({\it h})\equiv 0$;
precisely, $\partial_{\it v}{\it s}=\omega{\it h}$.
In terms of covariant derivative along the vector field $\frac{\partial}{\partial
z}$ we have system of differential equations
$$dY\,=\,\omega Y,\,\mbox{where}\quad \omega=B(z)dz.$$ Changing of the basis of sections
$$Y'=gY,\quad\mbox{for}\, g\in GL(2)$$ we also change the 1-form-valued $(2\times 2)$-matrix $\omega$:
$$\omega'=dg\cdot g^{-1}+g\cdot \omega\cdot g^{-1};$$ such transformations
are called gauge transformations. We suppose that all the singularities
of $B(z)$ are simple poles and we have the action of the monodromy: for
the loop $\gamma$ around some singular point the analitical
continuation along $\gamma$ gives $Y\rightarrow Y\cdot g_\gamma$, where
$g_\gamma$ is the monodromy matrix.
Suppose that $z=0$ is the singular point of $B(z)$ and $\gamma$ is the
apropriate element of fundamental group then the behavior of
the solution $Y(z)=(y_1(z), y_2(z))$ of our system
in the neighbourhood of the singularity at $z=0$ is described by the following\\
{\bf Fact.} The fundamental solution {\it Y(z)} admits the folowing presentation
$$Y(z)=U(z)\cdot z^A\cdot z^E,\,\mbox{where}\quad E:=\frac{1}{2\pi\sqrt{-1}}\log g_\gamma$$
and $A:=diag(\sigma_0, \tau_0)$ where $\sigma_0$ and $\tau_0$ are the
exponents of the components $y_1(z)$ and $y_2(z)$ of the solution in the neighbourhood of
$z=0$:
$$\mbox{exponent of}\,\,y(z):=\,
sup\{k\in\mathbb{Z}|\,\forall s<k\quad\frac{y(z)}{|z|^s}\rightarrow 0,\,z\rightarrow 0\};$$
the matrix $U(z)$ is holomorphic and invertible in the neighbourhood of
$z=0$.
\\
Consider the gauge transformation of special type:
$$ g_0:=U(z)\left(\begin{array}{cc}
    z & 0 \\
    0 & 1
    \end{array}\right)U(z)^{-1}$$
then its action $$g_0\cdot Y(z)\,=\,U(z)z^{A'}z^E,\,\mbox{where}\quad
A'=\left(\begin{array}{cc}
    \sigma_0 & 0 \\
    0 & \tau_0
    \end{array}\right)\,+\,
\left(\begin{array}{cc}
    1 & 0 \\
    0 & 0
    \end{array}\right)$$ shifts the exponents of the solution.
For abitrary simple pole $x_i$ of $B(z)$ ---
$$g_i:=U(z-x_i)\left(\begin{array}{cc}
    z-x_i & 0 \\
    0 & 1
    \end{array}\right)U(z-x_i)^{-1}\quad\mbox{or}\quad
U(z-x_i)\left(\begin{array}{cc}
    (z-x_i)^{-1} & 0 \\
    0 & 1
    \end{array}\right)U(z-x_i)^{-1}.$$
In this sense we may understand the
modifications of pairs $(\mathcal{L},\nabla)$ as singular gauge transformations and we have to check that
they do not change our system but reparametrise it.
\\
Suppose we have general linear differential equation on  the Riemann sphere $\mathbb{P}^1$ of the order
two which
have three simple poles at ${\it x}_{\rm 1}, {\it x}_{\rm 2}, {\it x}_{\rm 3}\in\mathbb{P}^1$.
Then the set of all its solutions may be encoded in the
$$\mbox{Riemann scheme of this equation}\qquad
    \left(\begin{array}{ccc}
    {\it x}_{\rm 1} & {\it x}_{\rm 2} & {\it x}_{\rm 3}\\
    \sigma _{\rm 1} & \sigma _{\rm 2} & \sigma _{\rm 3}\\
    \tau _{\rm 1} & \tau _{\rm 2} & \tau _{\rm 3}
    \end{array}\right).$$
$\{\sigma _{\it i}, \tau _{\it i}\}$ are the exponents of the solutions
at $\{{\it x}_{\it i}\}$ respectively; otherwise, if ${\it y}_1, {\it y}_2$
are the independent solutions at $z=a$ then
$${\it y}_1\sim({\it x-a})^{\sigma_{\it a}},\qquad
{\it y}_2\sim({\it x-a})^{\tau_{\it a}}.$$
They satisfy the Fuchs relation
$$ \sum _{\rm i=1}^{\rm 3}(\sigma _{\it i}+\tau _{\it i})~=~1;$$ it is
analogous to the determinant isomorphism $\phi$ from the third section and
it means that we handle with the bundle with trivial determinant on
$\mathbb{P}^1$.\\
It is well known that the differential equation is uniquely determined by its Riemann
scheme:
$$\frac{d^2 y}{dz^2}+\left(\sum_{i=1}^3\frac{1-\sigma _i-\tau _i}{z-x_n}\right)\frac{dy}{dz}+
\left(\sum^3_{i=1}\frac{\sigma _i\tau _i(x_n-x_{n+1})(x_n-x_{n+2})}{z-x_n}\right)\frac{y}{(z-x_1)(z-x_2)(z-x_3)}=0,$$
where $x_4=x_1~{\rm and}~x_5=x_2$.
\\
The fact is that every such equation one may reduce to the {\bf hypergeometric
equation}
$$z(1 - z)\frac{d^2 y}{dz^{2}} + [c - (a+b+1)z]\frac{dy}{dz} - ab\cdot y = 0.$$
$a, b, c \in\mathbb{C}\smallsetminus\{-1, -2,\ldots\}$ are the parameters of the
equation; the condition on them has the same meaning as the
eigenvalues-condition from the third section.
One can suppose that the singularities
$\{x_1, x_2, x_3\} = \{0, 1, \infty\}$;
on the language of the Riemann schemes it means we can reduce the
Riemann scheme to the special case
$$
 \left(\begin{array}{ccc}
    0 & 1 & \infty\\
    \sigma _{\rm 0} & \sigma _{\rm 1} & \sigma _\infty\\
    \tau _{\rm 0} & \tau _{\rm 1} & \tau _\infty
    \end{array}\right)
={\it z}^{\sigma _0}({\it z}-1)^{\sigma _1}
 \left(\begin{array}{ccc}
    0 & 1 & \infty\\
     0 & 0 & \sigma _\infty+\sigma _0+\sigma _1\\
    \tau _{\rm 0}-\sigma _{\rm 0} & \tau _{\rm 1}-\sigma _1 &
\tau _\infty+\sigma _0+\sigma_1
    \end{array}\right);
$$
in terms of previous chapters we perform
$$\left\{\begin{array}{ccc}
                \lambda_0 & \lambda_1  & \lambda_\infty\\
                -\lambda_0 & -\lambda_1 & -\lambda_\infty
              \end{array}\right\}
\longrightarrow
\left\{\begin{array}{ccc}
                0 & 0 & \lambda_0+\lambda_1+\lambda_\infty\\
                -2\lambda_0 & -2\lambda_1 & -\lambda_0-\lambda_1-\lambda_\infty
              \end{array}\right\}.$$
Further we compare our
calculations of $\widehat{\it C}_3$-group from the previous chapter with
classical calculations of K.-F. Gau$\ss$ and E. Kummer.
\\
A solution with the exponent 0 at $z=0$ of the equation is the hypergeometric
function
$$F(\alpha, \beta, \gamma~|~z) = \sum_{\it n}
\frac{(\alpha)_{\it n}(\beta)_{\it n}}{(\gamma)_{\it n}}\cdot\frac{z^n}{n!}\quad,$$
$$\mbox{where}\quad({\it a})_{\it n}:=\frac{\Gamma(a+n)}{\Gamma(a)}
\quad\mbox{and}\quad\Gamma(x)\, \mbox{is Euler's gamma-function}.$$
Further we investigate the symmetries of the equation and present
them in the form of relations between hypergeometric functions with
different parameters; then we analyse these relations in the sense of our
result of the previous sections.
\\
Immediately one can notice that this solution admits the obvious symmetry
$\beta\rightleftarrows\alpha$.
Of course the parameters $\alpha, \beta, \gamma$ determine the charachteristic
exponents at $0, 1, \infty$ by the rule
$$\left(\begin{array}{ccc}
    0 & 1 & \infty\\
    0 & 0 & \alpha\\
    1-\gamma & \gamma-\alpha-\beta & \beta
    \end{array}\right).$$
That means that we have two independent solutions of our equation with the
exponents \\ $\{0, 1-\gamma\}$ at 0
$${\it y}^{\it hol}_{0}=F(\alpha, \beta, \gamma~|~z), \qquad
{\it y}^{1-{\it \gamma}}_{0}={\it z}^{1-\gamma}F(\alpha+1-\gamma, \beta+1-\gamma, 2-\gamma~|~z);$$
the first is holomorphic (exponent 0) and the other has a monodromy (exponent
1-$\gamma$).
Permuting $0, 1, {\rm and}~ \infty$ by M\"{o}bius transformations:
$$(0~1): z\rightarrow 1 - z, \quad (0~\infty): z\rightarrow\frac{1}{z},\quad
\mbox{and}\quad (1~\infty): z\rightarrow\frac{z}{z -1}$$ we get
solutions at the other points
$${\it y}^{\it hol}_{1}=F(\alpha, \beta, \alpha+\beta+1-\gamma~|~1-z), \qquad {\it f}^{\it \gamma-\alpha-\beta}_{1}=(1-{\it z})^{\gamma-\alpha-\beta}F(\gamma-\alpha, \gamma-\beta, \gamma-\alpha-\beta+1~|~1-z),$$
$${\it y}^{\alpha}_{\infty}=z^{-\alpha}F(\alpha, \alpha+1-\gamma, \alpha+1-\beta~|~\frac{1}{z}), \qquad {\it f}^{-\beta}_{\infty}={\it z}^{-\beta}F(\beta, \beta+1-\gamma, \beta+1-\alpha~|~\frac{1}{z}).$$
From this one can deduce all 24 Kummer's solutions ([2]).
\\
Let us consider the action of modifications on these solutions without adding 1-form
$\omega_{ij}$ but we do not change symbol of modification.
$$(\uparrow\downarrow)_{1\infty}:
 \left(\begin{array}{ccc}
    0 & 1 & \infty\\
    0 & 0 & \alpha\\
    1-\gamma & \gamma-\alpha-\beta & \beta
    \end{array}\right)
\longrightarrow
 \left(\begin{array}{ccc}
    0 & 1 & \infty\\
    0 & 0 & \alpha+1\\
    1-\gamma & \gamma-\alpha-\beta-1 & \beta
    \end{array}\right)
\qquad{\rm i.e.}\quad
\alpha\rightarrow \alpha+1.$$
This shift of $\alpha$ does not change our hypergeometric equation.
In other words the apropriate Riemann's schemes are equivalent:
$$
 \left(\begin{array}{ccc}
    0 & 1 & \infty\\
    0 & 0 & \alpha\\
    1-\gamma & \gamma-\alpha-\beta & \beta
    \end{array}\right)
\approx
 \left(\begin{array}{ccc}
    0 & 1 & \infty\\
    0 & 0 & \alpha+1\\
    1-\gamma & \gamma-\alpha-\beta-1 & \beta
    \end{array}\right).
$$
Actually
$$ \alpha{\it z}^{\alpha-1}F(\alpha+1, \beta, \gamma) = \frac{\partial}{\partial{\it z}}z^\alpha
F(\alpha, \beta, \gamma),\quad\mbox{hence,}
\quad F(\alpha+1)=F(\alpha)+\frac{\it z}{\alpha}\frac{\partial}{\partial{\it z}}F(\alpha);$$
and for $\alpha':=\alpha+1-\gamma$ we have
$$\alpha'{\it z}^{\alpha'-1}F(\alpha'+1)=\alpha'{\it z}^{\it \alpha-\gamma}F(\alpha')+{\it z}^{\alpha'}F'(\alpha'). $$
So for another solution $F_2(\alpha'):={\it z}^{1-\gamma}F(\alpha+1-\gamma)$
$$F_2(\alpha'+1)=\frac{\alpha}{\alpha'}F_2(\alpha')+\frac{\it z}{\alpha'}F'_2(\alpha')$$
and finally
$$
\left\{\begin{array}{c}
F(\alpha+1, \beta, \gamma\,|{\it z})\\
{\it z}^{1-\gamma}F(\alpha+1-\gamma+1, \beta+1-\gamma, 2-\gamma,|{\it z})
\end{array}\right\}~=
\left\{\begin{array}{c}
{\displaystyle F(\alpha, \beta, \gamma\,|{\it z})+\frac{\it z}{\alpha}\,\frac{\partial}{\partial{\it z}}F(\alpha, \beta, \gamma\,|{\it z})}\\
 \\
{\displaystyle\frac{\alpha}{\alpha+1-\gamma}F_2+
\frac{\it z}{\alpha+1-\gamma}\,\frac{\partial}{\partial{\it z}}F_2}
\end{array}\right\};$$
it is called the Gau$\ss$\, relation.
\\
So one can express ${\it y}^\cdot _{1, \infty}(\alpha-1)$ in terms of  ${\it y}^\cdot _{1, \infty}(\alpha)$
and $\frac{\partial}{\partial z}{\it y}^\cdot _{1, \infty}(\alpha)$ and make
suitable reparametrization of the hypergeometric equation. As usual it is
the affine symmetry of the equation.
\\
Therefore the modifications together with the 24 Kummer's symmetries, assosiated with
the M\"obious transformations and with obvious symmetry
$$\sigma^\infty\,:\quad\alpha\rightleftarrows\beta$$
produce all discrete symmetries of the differential
equation on $\mathbb{P}^1$ of the order two which
have three simple poles at ${\it x}_{\rm 1}, {\it x}_{\rm 2}, {\it x}_{\rm 3}\in\mathbb{P}^1$.
These discrete symmetries were just studied in classical works of K.
Gau$\ss$ and E. Kummer. One can check their apropriate relations with the help
of the suitable $\widehat{C}_3$ Dynkin diagram:
$$(\uparrow\downarrow)^{+}_0\circ\sigma^0\quad\Longrightarrow\quad
(0\,1)\quad\longleftrightarrow\quad(1\,\infty)\quad\Longleftarrow\quad
(\uparrow\downarrow)^{+}_{\infty}\circ\sigma^{\infty}$$
\\
Let us explain this diagram. We use a pair of modifications at the same
point and do not add 1-form $\omega_{ij}$. The illustration of the symbols in terms of the Riemann scheme is the
following.

$$
(\uparrow\downarrow)^+_0~:~
\left(\begin{array}{ccc}
    0 & 1 & \infty\\
    0 & 0 & \alpha\\
    1-\gamma & \gamma-\alpha-\beta & \beta
    \end{array}\right)
\longrightarrow
\left(\begin{array}{ccc}
    0 & 1 & \infty\\
    0+1 & 0 & \alpha\\
    1-\gamma-1 & \gamma-\alpha-\beta & \beta
    \end{array}\right)
=
$$
$$={\it z}
\left(\begin{array}{ccc}
    0 & 1 & \infty\\
    0 & 0 & \alpha+1\\
    1-\gamma-2 & \gamma-\alpha-\beta & \beta+1
    \end{array}\right),
\quad\mbox{ i.e.}\,
\left\{\begin{array}{c}
\alpha\longrightarrow \alpha+1 \\
\beta\longrightarrow \beta+1 \\
\gamma\longrightarrow \gamma+2
\end{array}\right\};
$$
\\
$$
(\uparrow\downarrow)^-_\infty~:~
\left(\begin{array}{ccc}
    0 & 1 & \infty\\
    0 & 0 & \alpha\\
    1-\gamma & \gamma-\alpha-\beta & \beta
    \end{array}\right)
\longrightarrow
\left(\begin{array}{ccc}
    0 & 1 & \infty\\
    0 & 0 & \alpha+1\\
    1-\gamma & \gamma-\alpha-\beta & \beta-1
    \end{array}\right),
\quad\mbox{i.e.}\,
\left\{\begin{array}{c}
\alpha\longrightarrow \alpha+1 \\
\beta\longrightarrow \beta-1 \\
\end{array}\right\}.
$$
One can easily check these Coxeter relations in the same
way using the Kummer and Gau$\ss$ relations between corresponding solutions.

\section{Another example: W($\widehat{\it C}_4$)-symmetries of\\ the Heun equation}

Next step after hypergeometric equation is the Heun equation. Precisely, it
can be shown that any Fuchsian equation of the second order with four
singularities can be reduced to {\bf Heun's equation}:

$$\frac{{\it d}^2{\it y}}{{\it dz}^2}+\left[\frac{\gamma}{\it z}+\frac{\delta}{{\it z}-1}+
\frac{\epsilon}{\it z-a}\right]\frac{\it dy}{\it dz}+\frac{\alpha\beta\,{\it z-q}}
{{\it z}({\it z}-1)({\it z-a})}{\it y}=0; $$ the Fuchs relation is
$$\alpha+\beta-\gamma-\delta+1-\epsilon=0.$$
\\
We assume that the singularities are $\{0, 1, {\it a}, \infty\}$ and the Riemann scheme is
$$\left(\left.\begin{array}{cccc}
    0 & 1 & {\it a} & \infty\\
    0 & 0 & 0 & \alpha\\
    1-\gamma & 1-\delta & 1-\epsilon & \beta
    \end{array}\right|
\begin{array}{c}
     {\it q}\\
\end{array}\right),
$$
where {\it q} is the auxiliary parameter; the action of linear
transformations on the scheme is analogous to the previous case of
hypergeometric equation:
$$
\left(\left.\begin{array}{cccc}
    0 & 1 & {\it a} & \infty\\
    \sigma_0 & \sigma_1 & \sigma_{\it a} & \sigma_\infty\\
    \tau_0 & \tau_1 & \tau_{\it a} & \tau_\infty
    \end{array}\right|
\begin{array}{c}
     {\it q}\\
\end{array}\right)\quad=
{\it z}^{\sigma_0}({\it z}-1)^{\sigma_1}({\it z-a})^{\sigma_{\it a}}\,\times$$

$$\times\left(\left.\begin{array}{cccc}
    0 & 1 & {\it a} & \infty\\
    0 & 0 & 0 & \sigma_\infty+\sigma_0+\sigma_1+\sigma_{\it a}\\
    \tau_0-\sigma_0 & \tau_1-\sigma_1 & \tau_{\it a}-\sigma_{\it a} &
\tau_\infty+\sigma_0+\sigma_1+\sigma_{\it a}
    \end{array}\right|
\begin{array}{c}
     {\it q}'\\
\end{array}\right).$$
It should be noted that in the case of four singularities there is no
one-to-one correspondence between the Riemann schems and Fuchsian equations
that is why we need the auxillary parameter {\it q} and every transformation of the
equation acts it; the transformation of the auxilliary parameter must ascertained
by explicit calculation. The apropriate computation will be needed in the next section and
we shall make it with the help of geometric interpretation of the Fuchsian differential
equation with four singularities.

Then as usual we consider the following linear
transformations of projective line
$${\it z}\longrightarrow1-{\it z},\quad{\it z}\longrightarrow\frac{1}{\it z},\quad
{\it z}\longrightarrow\frac{\it a}{\it z},\quad{\it z}\longrightarrow{\it z-a},\quad
{\it z}\longrightarrow\frac{1-{\it a}}{\it z-a}$$
and thier compositions; for each point there are six such expressions, for example, six possible values of {\it a} are
$$1-{\it a},\quad\frac{1}{\it a},\quad\frac{1}{1-{\it a}},\quad\frac{\it a}{{\it a}-1},
\quad\frac{{\it a}-1}{\it a}.$$ There are 24 such transpositions at all and they just have the
meaning of permutations of four singular points.
\\
Following K. Heun we denote
$${\it y}^{\it hol}_0\,=\,F({\it a}\,|\,\alpha, \beta, \gamma, \delta|\,{\it z})$$ the
holomorphic solution at $z=0$; it is also admits the symmetry
$\sigma^\infty:\alpha\rightleftarrows\beta.$
The nonholomorphic is
$${\it y}^{1-\gamma}_0\,=\,
{\it z}^{1-\gamma}F({\it a}\,|\alpha+1-\gamma, \beta+1-\gamma, \gamma, \delta|\,{\it z});$$
so let us consider the action of linear transformations on it. We can get local
solutions at other singularities
$${\it y}^{\it hol}_1\,=\,F(1-{\it a}\,|\alpha, \beta, \gamma, \delta|\,1-{\it z}),\quad
{\it y}^{1-\delta}_1\,=\,({\it z}-1)^{1-\delta}({\it z-a})^\alpha F({\it a}\,|\alpha+1-\delta,
\gamma+1-\beta, \delta, 2-\gamma\,|\frac{{\it a}({\it z}-1)}{\it z-a});$$

$${\it y}^{\it hol}_{\it a}\,=\,F(\frac{1-{\it a}}{\it a}\,|\alpha, \beta, \alpha+\beta-\gamma-\delta+1,
\delta, 2-\gamma\,|\,\frac{\it a-x}{\it a}),$$
$${\it y}^{1-\epsilon}_{\it a}\,=\,({\it x}-1)^
{\gamma+\delta-\alpha-\beta}F(\frac{1}{1-{\it a}}\,|\,\gamma+\delta-\beta, \gamma+\delta-\alpha,
\delta, \gamma+\delta-\alpha-\beta+1\,|\,\frac{{\it x}-1}{\it x-a});$$

$${\it y}^\alpha_\infty\,=\,{\it z}^\alpha F(\frac{1}{\it a}\,|\alpha, \alpha+1-\gamma,
\beta+1-\alpha, \delta\,|\,\frac{1}{\it z}),\quad{\it y}^\beta_\infty\,=\,{\it z}^\beta
F(\frac{1}{\it a}\,|\beta, \beta+1-\gamma, \beta+1-\alpha, \delta\,|\,\frac{1}{\it z})$$
and all 192 expressions for these solutions. In such way we calculate the
action of the finite part of our group on the solutions and together with
the obvious symmetry
$$\sigma^\infty\,:\quad\alpha\rightleftarrows\beta$$
we describe the finite part of our group of order
$$|W(C_4)|\,=\,|\left(\mathbb{Z}/2\mathbb{Z}\right)^4\rtimes\mathfrak{S}_4|\,=\,192\times 2.$$
\\
To check the action of the affine part we need a relation on shifted and non-shifted
functions ${\it F}({\it a}\,|\,{\it z})$ analogous to the Gau\ss\, relation for the
hypergeometric function; it is the folowing.
$$[(\epsilon-1)\,-\,\frac{\alpha\beta}{\gamma}q\,]\,
F({\it a, q'}\,|\,\alpha, \beta, \gamma+1, \delta+1\,|\,{\it z})\,=$$
$$=\,(\epsilon-1)F({\it a, q}\,|\,\alpha, \beta, \gamma, \delta\,|\,{\it z})\,+\,(\it z-a)
\frac{\partial}{\partial{\it z}}F({\it a, q}\,|\,\alpha, \beta, \gamma, \delta\,|\,{\it z}),$$
where the modified auxiliary parameter is
$${\it q'}\,=\,q\,+\,{\it a}\frac{\gamma+\delta}{\alpha\beta}\,-\,\frac{\gamma}{\alpha\beta};$$
we give the explanation of the modification auxiliary parameter in the next section using
beautiful geometric interpretation and hamiltonian presentation of the isomonodromic deformation of
the Heun equation. In hamiltonian presentation auxilary parameter plays the role of the inverse
to the moment variable.
\\
The above transformation relates the solution with shifted parameters $$\gamma\rightarrow\gamma+1,\quad
\delta\rightarrow\delta+1,\quad\epsilon\rightarrow\epsilon-1$$ to the linear
combination of non-shifted solution and its derivative. So one can
reparametrise the eqation and assure that the affine action of the {\it
W($\widehat{C}_4$)} does not change the equation. The simple computation is
the following; we have to look after the second solution:
$$\frac{\partial}{\partial{\it z}}{\it z}^{1-\gamma}\,F(\alpha+1-\gamma,
\beta+1-\gamma, \gamma, \delta)\,=$$
$$=\,(1-\gamma){\it z}^{-\gamma}F(\alpha+1-\gamma,
\beta+1-\gamma, \gamma, \delta)\,+\,{\it z}^{1-\gamma}F'(\alpha+1-\gamma,
\beta+1-\gamma, \gamma, \delta)$$ so we substitute it to the above relation
and finally get a reparametrisation for a pair of solutions at $z=0$
$$
\left\{\begin{array}{c}
F({\it q'}\,|\alpha, \beta, \gamma+1, \delta+1\,|\,{\it z})\\
 \\
{\it z}^{1-\gamma}F({\it q'}\,|\alpha+1-\gamma-1, \beta+1-\gamma-1, \gamma+1, \delta+1|\,{\it z})
\end{array}\right\}~=~$$

$$\left\{\begin{array}{c}
{\displaystyle \frac{\epsilon-1}{\epsilon-1-\frac{\alpha\beta}{\gamma}\,{\it q}}
F({\it q}\,|\alpha, \beta, \gamma, \delta\,|\,{\it z})+
\frac{\it z-a}{\epsilon-1-\frac{\alpha\beta}{\gamma}\,{\it q}}\,\frac{\partial}{\partial{\it z}}
F({\it q}\,|\alpha, \beta, \gamma, \delta\,|\,{\it z})}\\
 \\
{\displaystyle\frac{\epsilon-1-(1-\frac{\it a}{\it z})(1-\gamma)}{\epsilon-1-\frac{\alpha\beta}{\gamma}\,{\it q}}
F_2\,+\,\frac{\it z-a}{\epsilon-1-\frac{\alpha\beta}{\gamma}\,{\it q}}\,
\frac{\partial}{\partial{\it z}}F_2}
\end{array}\right\},$$ \\ where
$F_2\,=\,{\it z}^{1-\gamma}F({\it q}\,|\alpha+1-\gamma, \beta+1-\gamma, \gamma, \delta|{\it
z})$. If we combine it with the action of permutational part we get the
reparametrisations for local solutions at other points.
\\
Just like in the case of hypergeometric equation one can combine
192 permutational and translational Heun's relations with the transposition
$\sigma^\infty\,:\,\alpha\rightleftarrows\beta$ and get
$W(\widehat{\it C}_4)$ group of symmetries.

\section{The isomonodromic deformation of the Heun\\ equation: sixth Painlev\'e equation}

And finally let us consider the equation of isomonodromic deformation of the fuchsian
differential equation of order two with four singularities on the projective line.
\\
There is another way of considering our problem;
we say that an algebraic differential equation of order two satisfies the
Painlev\'e property if it is free from movable branch points.
Sixth Painlev\'e equation {\it P}$_{\it VI}$ is the general differential equation of
order two on $\mathbb{P}^1$ with at most four regular singularities and without
movable branch points; as
usual we may assume that singular points are 0, 1, ${\it t}, \infty.\,$
{\it P}$_{\it VI}$ is the following.
$$\frac{d^2x}{dt^2}~=~\frac{1}{2}\left(\frac{1}{x}+\frac{1}{x-1}+\frac{1}{x-t}\right)
\left(\frac{dx}{dt}\right)^2-
\left(\frac{1}{t}+\frac{1}{t-1}+\frac{1}{x-t}\right)\frac{dx}{dt}+$$
$$\frac{x(x-1)(x-t)}{t^2(t-1)^2}\left(\alpha-\beta\frac{t}{x^2}+\gamma\frac{t-1}{(x-1)^2}+(\frac{1}{2}-\delta)
\frac{t(t-1)}{(x-t)^2}\right).$$
Time variable is the double ratio of the singular points: ${\it t}=[0,1,{\it t},
\infty]$.
The parameters $\alpha, \beta, \gamma, \delta$ are represent the eigenvalues
$\lambda_0, \lambda_1, \lambda_{\it t}, \lambda_\infty$
of the residues of logarithmic connection in the following way.
$$\alpha=\frac{1}{2}\lambda_\infty^2,\quad \beta=\frac{1}{2}\lambda_0^2,\quad
\gamma=\frac{1}{2}\lambda_1^2,\quad \delta=\frac{1}{2}\lambda_{\it t}^2.$$
For our purposes the Hamiltonian form of {\it P}$_{\it VI}$ is more
suitable; it is the following.
$$
\left\{\begin{array}{l}
{\displaystyle
\frac{\it dx}{\it dt}=\frac{\it \partial H}{\it \partial p},}\\
\\
{\displaystyle
\frac{\it dp}{\it dt}=-\frac{\it \partial H}{\it \partial x},}\\
\end{array}\right.
$$
\\
with the Hamiltonian
$${\it H}~=~\frac{1}{{\it t(t}-1)}[{\it x(x}-1)({\it x-t)p}^2-\{\lambda_0{\it
(x}-1)({\it x-t)}+$$
$$+\lambda_1{\it x(x-t)}+(\lambda_{\it t}-1){\it x(x}-1)\}{\it p}+\lambda{\it (x-t)}],$$
where $\lambda=\frac{1}{4}[(\lambda_0+\lambda_1+\lambda_{\it t}-1)^2-\lambda_\infty^2]$.
The group of symmetries is isomorphic to $W(\widehat{\it F}_4)$. Its structure is
similar to the one in general case and in the case of hypergeometric equation.

Moduli space $\mathcal{M}_{\rm 4}$ admits a group $W(\widehat{\it D}_4)$ of
transformations, that is generated only by short shifts or only by pairs of
modifications at distinct points. The consideration of this subgroup of
$W(\widehat{\it C}_4)$ is quite natural in the following geometrical
interpretation ([1]). The singularities of our connection is at the support
of the module $\mathfrak{M}:=0+1+{\it t}+\infty$ on $\mathbb{P}^1$. The initial data space of
the isomonodromic system {\it P}$_{\it VI}$ is isomorphic to the noncompact
complex surface $$\{({\it x, p})~|~({\it x, p})^{\it up}:\mathcal{O}\oplus\mathcal{T}(-{\mathfrak M})
\rightarrow\mathcal{O}\oplus\mathcal{O}(-\infty)\};$$ locally it is
isomorphic to $\mathcal{T}^*(\mathbb{P}^1-{\it S})$; the structure of this surface was thoroughly studied
in [6, 7] (see also [2]) and [1]. Then moment variable {\it
p} has the meaning of the direction of the upper modification of bundle
$(\mathcal{O}\oplus\mathcal{T}(-\mathfrak{M}))$ at point {\it x}; in the above
notation the hamiltonian
{\it H} corresponds to $({\it x, p})^{\it up}:\,\mathcal{O}\oplus\mathcal{T}(-\mathfrak{M})
\longrightarrow\mathcal{O}\oplus\mathcal{O}(-{\it t})$ that is not important due to our symmetries.
Precisely, the action of the pairs of modifications at distinct points is obvious; it is the
following
$$\mathcal{O}\oplus\mathcal{O}(-\infty)\longrightarrow\mathcal{O}\oplus\mathcal{O}(-{\it x_i}),
\qquad{\it x_i}=0,\,1,\,{\it t}.$$
In matrix presentation
$$({\it x, p})^{\it up}~=~
\left(\begin{array}{cc}
\omega & \eta \\
1 & 0
\end{array}\right), $$ where
$\omega\,\in{\it Hom}(\mathcal{T}(-\mathfrak{M}), \mathcal{O})
\simeq\Omega(\mathfrak{M})$,  nd $\eta\,\in\Omega(\mathfrak{M}-\infty).$
\\
The determinant ${\it det} ({\it x, p})^{\it up}$ has a simple zero at {\it x},
so
$$\eta({\it x})\,=\,0,\qquad \omega({\it x})\,=\,{\it p\,dx}.$$
This construction is used in the work [21] of E. Sklyanin for separating the
variables in $sl(2)$-Schlesinger system with $n$ singularities.
\\
For example, the shifts of the parameters for a pair of modifications at
points {\rm 0} and {\rm 1} are
$$\quad{\it
p}\longrightarrow{\it p}+\triangle{\it p};\qquad\triangle{\it
p}~=~\frac{1}{{\it x}-1}-\frac{1}{\it x},\qquad \mbox{ nd}\quad{\it x}\quad
\mbox{does not changed};$$ these transformation preserves the hamiltonian:
$${\it
H}^{''}~=~\frac{1}{{\it t(t}-1)}[{\it x(x}-1)({\it x-t)(p+\triangle p)}^2-
\{(\lambda_0+\frac{1}{2}){\it (x}-1)({\it x-t)}+$$
$$+(\lambda_1-\frac{1}{2}){\it x(x-t)}+(\lambda_{\it t}-1){\it x(x}-1)\}
({\it p+\triangle p})+{\it k(x-t)}]~=~{\it H}+2{\it p}\frac{1}{2}({\it
x-t})+$$ $$+\frac{1}{4}\frac{\it x-t}{{\it
x(x}-1)}-\frac{1}{2}\left[\lambda_0 \frac{\it x-t}{\it x}+\lambda_1\frac{\it
x-t}{{\it x}-1}+ (\lambda_{\it t}-1)\right]-\frac{1}{2}({\it
x-t})-\frac{1}{2}\frac{1}{2} \frac{\it x-t}{{\it x(x}-1)}~=~{\it H}.$$
\\
These $\widehat{\it D}_4$-symmetries we can extend with the group
$\mathfrak{S}_3$ of automorphisms of the graph ${\it D}_4$ and get the
affine group $W(\widehat{\it F}_4)$ of symmetries of the {\it P}$_{\it VI}$
equation. The group $W(\widehat{\it F}_4)$ contains the group $W(\widehat{\it C}_4)$
that we get in the general case. However the algebraic stack $\mathcal{M}_4$ admits
unusually large group of symmetries because of the existance of exeptional
graph automorphisms.

\end{document}